%% file: main.tex
\documentclass[
    amsmath,
    amssymb,
    aps,
    twocolumn,
    superscriptaddress
]{revtex4-2}

\usepackage[utf8]{inputenc}
\usepackage{float}
\usepackage{graphicx}
\usepackage{color}
\usepackage[dvipsnames]{xcolor}
\graphicspath{ {./images/} }

\input{commandsNotation.tex}

\begin{document}
\title{Rigidity percolation in a random tensegrity via analytic graph theory}
\author{William Stephenson}
\affiliation{School of Physics, Georgia Institute of Technology, Atlanta, GA 30332}
\author{Vishal Sudhakar}
\affiliation{School of Physics, Georgia Institute of Technology, Atlanta, GA 30332}
\author{James McInerney}
\affiliation{Department of Physics, University of Michigan, Ann Arbor, MI 48109} 
\author{Michael Czajkowski}
\affiliation{School of Physics, Georgia Institute of Technology, Atlanta, GA 30332}
\author{D. Zeb Rocklin}
\affiliation{School of Physics, Georgia Institute of Technology, Atlanta, GA 30332}
\date{\today}

\begin{abstract}
Functional structures from across the engineered and biological world combine rigid elements such as bones and columns with flexible ones such as cables, fibers and membranes. These structures are known loosely as tensegrities, since these cable-like elements have the highly nonlinear property of supporting only extensile tension. 
Marginally rigid systems are of particular interest because the number of structural constraints permits both flexible deformation and the support of external loads. We present a model system in which tensegrity elements are added at random to a regular backbone. This system can be solved analytically via a directed graph theory, revealing a novel mechanical critical point generalizing that of Maxwell. We show that even the addition of a few cable-like elements fundamentally modifies the nature of this transition point, as well as the later transition to a fully rigid structure. Moreover, the tensegrity network displays a fundamentally new collective avalanche behavior, in which the addition of a single cable leads to the elimination of multiple floppy modes, a phenomenon that becomes dominant at the transition point. These phenomena have implications for systems with nonlinear mechanical constraints, from biopolymer networks to soft robots to jammed packings to origami sheets.
\end{abstract}

\maketitle

\section{Introduction}

Hooke's law, suitably generalized, accurately describes how a solid body possesses equal stiffness against infinitesimal compression and extension, whether that body is a microscopic particle or towering skyscraper. This paradigm of linear elasticity is relatively well-understood~\cite{landau1986theory} via patterns of stress and strain and periodic waves. 
Structures assembled from rigid elements can realize an elastic solid via a rigidity transition whereby it loses the ability to deform without energy cost.
The linearity of constraints imposed by rigid elements enables simple counting arguments due to Maxwell~\cite{maxwell1864calculation,calladine1978buckminster}, and hence are referred to as Maxwell counting, that determine this onset of rigidity in the bulk of a lattice of harmonic springs in terms of a critical coordination number. However, nonlinear elasticity presents a far richer and more challenging range of behavior, including buckling and wrinkling instabilities~\cite{cerda2002wrinkling,cerda2003geometry,davidovitch2011prototypical}, solitons~\cite{chen2014nonlinear}, and plasticity and fracture~\cite{long2021fracture}, and the onset of rigidity remains a difficult question.

A characteristic example of nonlinearity can be found in systems called \emph{tensegrities}, in which there are one-way constraints that can support either positive or negative tensions but not both ~\cite{roth1981tensegrity}.
Such structures are composed of three types of elements that direct forces along their axis: (i) harmonic springs that resist extension and compression, (ii) cables that exclusively resist extension, (iii) and struts that exclusively resist compression. These tensegrity structures can be found in environments ranging from the cellular level~\cite{ingber2014tensegrity} to macro-scale architecture~\cite{song2022form}. However, the nonlinear nature of the network elements, which arises from the directionality of the restoring forces, makes analytical computations difficult and generic tensegrities may be rigid only at \emph{second} order in the deformation of the structure or in the presence of prestress~\cite{connelly1992stability,connelly1996second}. Thus, tensegrities do not admit the simple Maxwell counting arguments for rigidity and the rigidity of random tensegrities has not been well explored.

For generic elastic networks at zero temperature, the onset of rigidity occurs at the \emph{rigidity percolation} transition, which is a special case of more generic percolation transitions that characterize phenomena from traffic jams~\cite{vivek2019cyberphysical, biham1992self} and the spread of infectious diseases~\cite{schulman2021things} to the flow of current through an electrical circuit. These examples possess scalar degrees of freedom that facilitate analysis, even in the presence of one-way interactions~\cite{domany1984equivalence}. The rigidity percolation transition is particularly difficult to characterize, even in the linear regime, due to the vectorial nature of the forces. In fact, this transition for random networks of harmonic springs admits both first- and second-order character \cite{ellenbroek2011rigidity}. Moreover, the Maxwell counting arguments provide only an approximation for the rigidity transition.

These difficulties make it challenging to get rigidity percolation results at all, even without considering nonlinearities such as those present in tensegrities. Our work shows that it is possible. Through the random additions of rods, cables, and struts to a finite-sized square lattice backbone, we build on the bipartite graph theory model of Ref.~\cite{ellenbroek2011rigidity}. This model system contains degrees of freedom consisting of shears to entire rows or columns of pores simultaneously, where the introduction of a rod (cable/strut) across a pore results in an equality (inequality) between the row and column modes to which the pore belongs.

\begin{figure*}[t]
    \centering
    \includegraphics[width=0.95\textwidth]{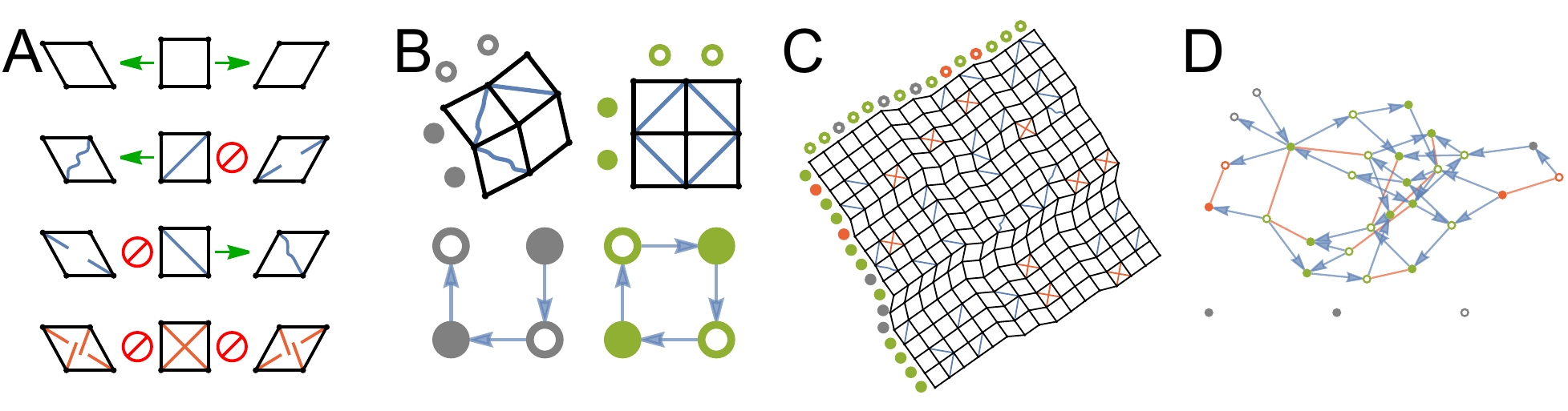}
    \caption{
    Systems of rigid rods acquire rigidity via the addition of random additional rods and cables, as captured via a graph theory.
    (A)  A single square pore can shear nonlinearly in either direction. The addition of a (blue) cable permits shear in one direction but not the other, depending on the orientation of the cable, while one or more (red) rigid rods rigidifies the pore.
    (B) The two columns and rows of a square of four pores (upper left) can undergo four separate shearing motions that persist even under the addition of three cables. The addition of a fourth cable (upper right) rigidifies the structure. This is represented as a graph in which each row (column) mode is a shaded (unshaded) node and each cable is a directed edge between the corresponding column and row. In the lower-left, the four  modes are weakly connected  but in the lower-right, the four modes have become one strongly connected cluster.
    (C) Our main object of study are structures that consist of large numbers of pores arranged in columns and rows with cables and rods added at random.
    (D) In the graph representation of (C), each cluster represents a distinct mode, and the coupling of modes beyond the column and row joined by an individual rod (red) comes from the emergence of a single giant cluster (green).}
    \label{fig:systemgrid}
\end{figure*}

The present work expands upon this mapping between rigidity and connectivity percolation by considering the rigidity of a square backbone with randomly added tensegrity elements. In contrast to springs, cables and struts map to one-way connections in a connectivity percolation problem that only rigidify pores in the real system when they form closed circuits in the graph. This method enables us to analytically identify and characterize the critical points at which the system begins to collectively support externally applied stresses and at which it becomes fully rigid, and we find that even a few cables are sufficient to shift and modify the character of these transitions relative to those observed in systems of purely rigid random elements. Moreover, we show that these systems are able to exhibit avalanches where a single tensegrity element eliminates a large number of degrees of freedom at once.

\section{A model random tensegrity and its graph-theoretical representation}

We consider a system that initially consists of particles lying on points of a Cartesian grid joined by inextensible rods. Within the square pores between the particles, cables and additional rods are added at random to connect particles to their next nearest neighbors (see \myfig{systemgrid}).

Prior to the addition of random elements, the bulk of the system is at the critically coordinated Maxwell point~\cite{pellegrino1986matrix,lubensky2015phonons,mao2018maxwell,Rocklin2020}, meaning that the number of rigid deformations of the structure is proportional to the length of the boundary. 
Convenient to this particular choice of model system, the nonlinear deformations are exactly known: shears of entire rows and columns (see \myfig{systemgrid}, B/C) \cite{ellenbroek2011rigidity}, such that a system with $\n$ rows and columns has a $2 \n$-dimensional nonlinear space of configurations, which includes a global rotation.
The shearing angle $\shear_{i,j}$ of a square pore in the $i^\textrm{th}$ column and $j^\textrm{th}$  row is then a combination of the shear on column $i$ and row $j$ (see \myfig{systemgrid}):

\begin{align}
    \shear_{i,j} = \col_i - \row_j.
\end{align}

This shearing has the effect of moving two particles on opposite sides of the pore closer and the other two further apart.
Consequently, a rigid rod placed across a pore connecting either pair of opposite points necessitates that $\shear_{i,j} = 0$. 
If instead a cable is placed, the connected pair of points would not be able to separate further from each other, imposing the constraint $\shear_{i,j} \ge 0$ or $\shear_{i,j} \le 0$, depending on the orientation of the cable.
Placing struts, structural elements that resist compression but not extension, across the pores has the same effect as cables of the opposite orientation. Consequently, our results apply equally to systems with random struts but for simplicity we only explicitly consider cables.

The addition of a rigid rod then combines a column mode and a row mode and reduces the space of configurations by one ($c_i = r_j$). In contrast, the addition of cable would rule out half of all possible configurations ($c_i \ge r_i$) but would not seem to reduce the \emph{dimensionality} of the space of configurations. However, \emph{multiple} cables can reduce the dimension of this space. Consider, in particular four cables placed in a $2\times 2$ square so that

\begin{align}
    \cOne \le r_1 \le \cTwo \le r_2 \le \cOne.
\end{align}

\noindent Prior to the addition of the fourth cable, the initial cables still allow ranges of choices for all four amplitudes and hence do not eliminate any floppy modes. However, once the fourth cable is placed, all four amplitudes must become equal, eliminating three deformations at once (See \myfig{systemgrid} B). 

In order to represent this set of physical relationships, we pass from our physical system to an abstracted graph system, where shear modes are represented by vertices of two types (row and column modes) in a bipartite graph~\cite{ellenbroek2011rigidity}. We then draw edges between nodes such that a rigid rod between column $i$ and row $j$ becomes an edge between the $i^\textrm{th}$ column vertex in the graph and the $j^\textrm{th}$ row vertex, indicating that these two amplitudes must be equal. 
In contrast, cables (and struts) are represented by directed edges which indicate that the starting vertex must have an amplitude at least that of the ending vertex.

We then invoke the notion of a \emph{strongly connected cluster} within the graph (See \myfig{systemgrid} D), consisting of sets of nodes 
in which every node are reachable from every other node (while going in only the allowed direction along directed edges)~(\cite{newman2018networks} Ch.~6.4). Each strongly connected cluster represents a distinct nonlinear mode, we therefore determine the number of deformation modes of the random tensegrity structure by examining the statistics of random directed graphs. We find that by adding rods and cables to eliminate deformation modes, the square lattice undergoes a phase transition where many of its pores begin to rigidify in a short window.

\section{Emergence of collective rigidity}

While placing a cable or rod in a pore respectively restricts or prevents that pore's \emph{individual} shearing, it is not obvious when and how these elements \emph{collectively} restrict the system's movement and support external loads. As the graph theory introduced in the previous section reveals, an individual pore is rigid precisely when its two associated deformation modes belong to the same \emph{cluster}, or set of nodes which can all be reached from one another. Here, we present a mean-field theory demonstrating that this collective rigidity is dominated by a giant cluster which couples an extensive fraction $f_b$ of the deformation modes. Consequently, a fraction $f_b^2$ of the pores in large systems are rigid.

In order to develop an analytic theory, we shall apply a mean-field approach in which the probabilities that different nodes are in a cluster are not correlated, inspired by approaches for undirected~(\cite{newman2018networks} Ch.~12) and directed~\cite{karp1990transitive} graphs. In any graph, in order for a vertex to be in a strongly connected cluster, it must be reachable from (``downstream'') and be able to reach (``upstream'') all of the other vertices in that cluster. 
We denote the fraction of vertices upstream from the giant cluster as $\fu$ and the fraction downstream as $\fd$
. Also important is the fraction of vertices that are either upstream or downstream from the giant cluster, $\fe$. Because of symmetry in the orientation of directed bonds, the expected values of $\fu$ and $\fd$ are the same. By the principle of inclusion-exclusion, we also get $\fb = \fu + \fd - \fe = 2\fu - \fe$. 

In order to calculate $\fu$ and $\fe$ to find $\fb$, we create two self-consistent mean field equations. For $\fu$, the probability that a vertex $\vertexi$ is \emph{not} upstream from the giant cluster through a vertex $\vertexj$ is the probability that there is no upstream bond from $\vertexi$ to $\vertexj$ plus the probability that such a bond does exist, but $\vertexj$ is not upstream from the giant cluster. For $\vertexi$ to not be upstream from the giant cluster, this condition must be true for all $\n$ possible choices of $\vertexj$. As nothing has been specified about $\vertexi$, the probability that it is upstream from the giant is $\fu$, the probability that any given node is upstream from the giant. Using these relations, we obtain the self-consistency condition for $\fu$ in the large-$\n$ limit,

\begin{align}\label{eq:fuTranscendental}
1 - \fu = \left[1-\frac{1}{n}\left(\frac{\cOne}{2}+\cTwo\right)\fu\right]^n \approx e^{-\left(\cOne/2 + \cTwo\right) \fu},
\end{align}

\noindent where $\cOne$ and $\cTwo$ are the directed and undirected coordination numbers respectively, defined as the expected number of cables ($\cOne$) or rods ($\cTwo$) per row or column of the lattice. 

Similarly, for a vertex to not be either upstream or downstream from the giant cluster, it must not have a directed edge to any upstream vertex, a directed edge from any downstream vertex, or an undirected edge to either an upstream or downstream vertex. As found in Methods:

\begin{align}
\label{eq:feTranscendental}
1-\fe = \left[1-\frac{1}{n}\left(\cOne\fu+\cTwo\fe\right)\right]^n \approx e^{-(\cOne\fu + \cTwo\fe)}.
\end{align}

\begin{figure*}[t]
    \centering
    \includegraphics[width=2\columnwidth]{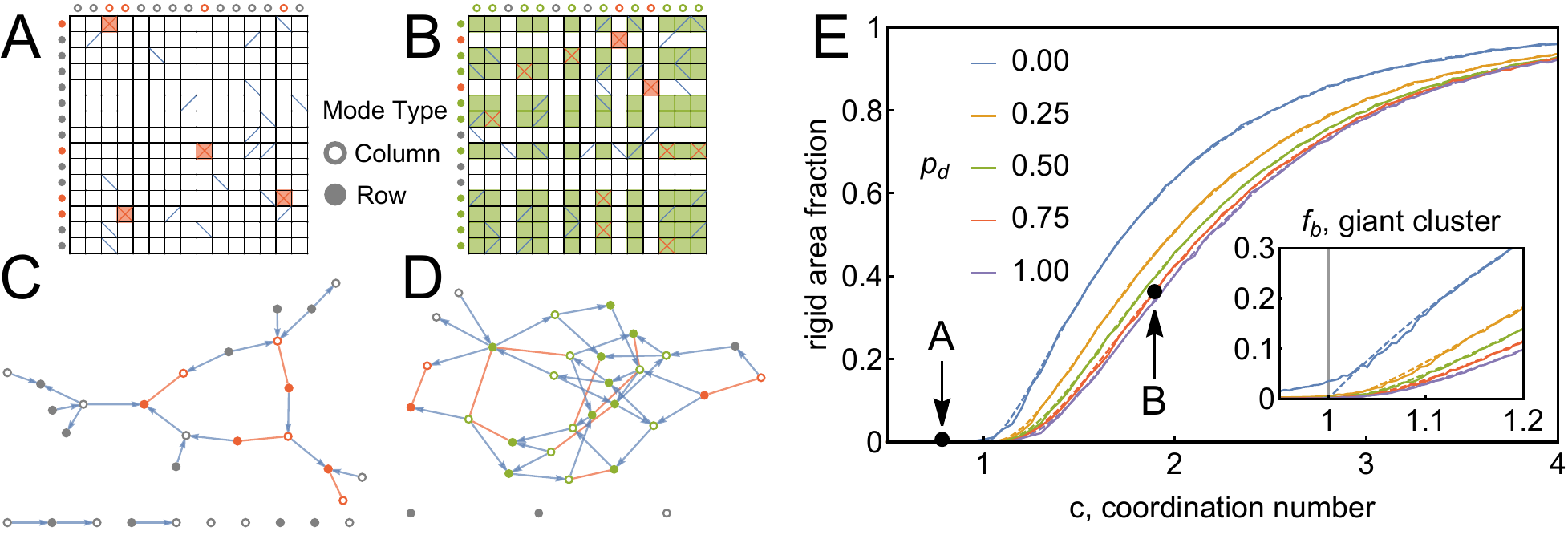}
    \caption{
   Following the addition of many rods and cables, a phase transition occurs, resulting in large rigid areas reflecting an extensive cluster of vertices in the graph theory representation. 
   As cables and rods are added at random, at low densities (A) they do not couple to one another and the only rigid pores are those with rods present (red). At high densities (B), the random elements collectively rigidify large portions (green) of the system,  including those with no local element. This is reflected in the corresponding graphs, which at low densities (C) do not have clusters of nodes that are strongly connected (other than the pairs joined by individual rods) whereas at high densities (D) a giant cluster emerges. (E) As the weighted number of elements per column, $\coord$, grows beyond the critical value of one, square pores become rigid. This growth is quadratic when purely rigid elements are added ($\pd = 0$), but softens to quartic growth for cables ($\pd = 1$) or a mixture of rods and cables. In the thermodynamic limit, the simulation results (solid lines, 1000 rows and columns) match closely to the theoretical predictions (dashed lines) derived from the fraction of nodes, $\fb$, in the giant cluster of the graph theory (inset). Values of $\coord$ corresponding to the systems in A, B are marked, though finite-size effects lead to deviations from the theory.}
    \label{fig:rigidAreaFrac}
\end{figure*}

While these self-consistency conditions cannot be solved analytically, it can be shown via Taylor Expansion that \myeq{fuTranscendental} [and therefore \myeq{feTranscendental}] begins to have nonzero solutions when the overall coordination number $\coord \equiv \cOne/2 + \cTwo$ is greater than 1. Before this transition at $\coord=1$, as in \myfig{rigidAreaFrac} C, no large clusters exist in the bipartite graph. Consequently, the pores in \myfig{rigidAreaFrac} A are rigid only if they contain a rigid element. After the transition, as in \myfig{rigidAreaFrac} B, the vast majority of rigid pores are not rigid because of a contained rod, but because their row and column shear modes both belong to the giant cluster in \myfig{rigidAreaFrac} D.

While this phase transition occurs at $\coord=1$ independently of the relative densities of rods and cables, the nature of the growth of the giant cluster beyond this point proves highly sensitive to this composition. Therefore, we define $\pd$, the weighted cable fraction (or the weighted probability that any bond in the graph is directed)

\begin{align}
\pd \equiv \frac{\cOne/2}{\cOne/2+\cTwo} = \frac{\cOne/2}{\coord}.
\end{align}

\noindent As shown in Methods, the growth of the giant cluster in the regime right after the transition is linear in the unique case of all rods ($\pd=0$), but generally quadratic in the presence of any cables ($\pd>0$). These two giant cluster growth cases result in quadratic and quartic growth of the rigid area fraction respectively, supported by the simulation results in \myfig{rigidAreaFrac} E. These simulation results show close quantitative agreement with the mean-field theory in all regimes for all cases except for the case of a system purely of rods very close to the transition point (inset).

As more elements are added, the rigid area fractions become qualitatively similar for all cable fractions. By the time that a system has several elements per column nearly all of its pores are rigid, despite the fact that nearly all of them are empty in the thermodynamic limit. In the next section, we discuss how the few remaining non-rigid areas are eliminated, resulting in the system becoming fully rigid.

\section{Transition to full rigidity}

As shown in the previous section, the emergence of collective rigidity eliminates certain modes of deformation, but it does not imply that the structure is fully rigid. Here, we examine the condition for full rigidity, in which there are no floppy modes remaining. Such a fully rigid system is fixed in its square reference shape and is capable of supporting stress across the entire lattice.

\begin{figure*}[t]
    \centering
    \includegraphics[width=2\columnwidth]{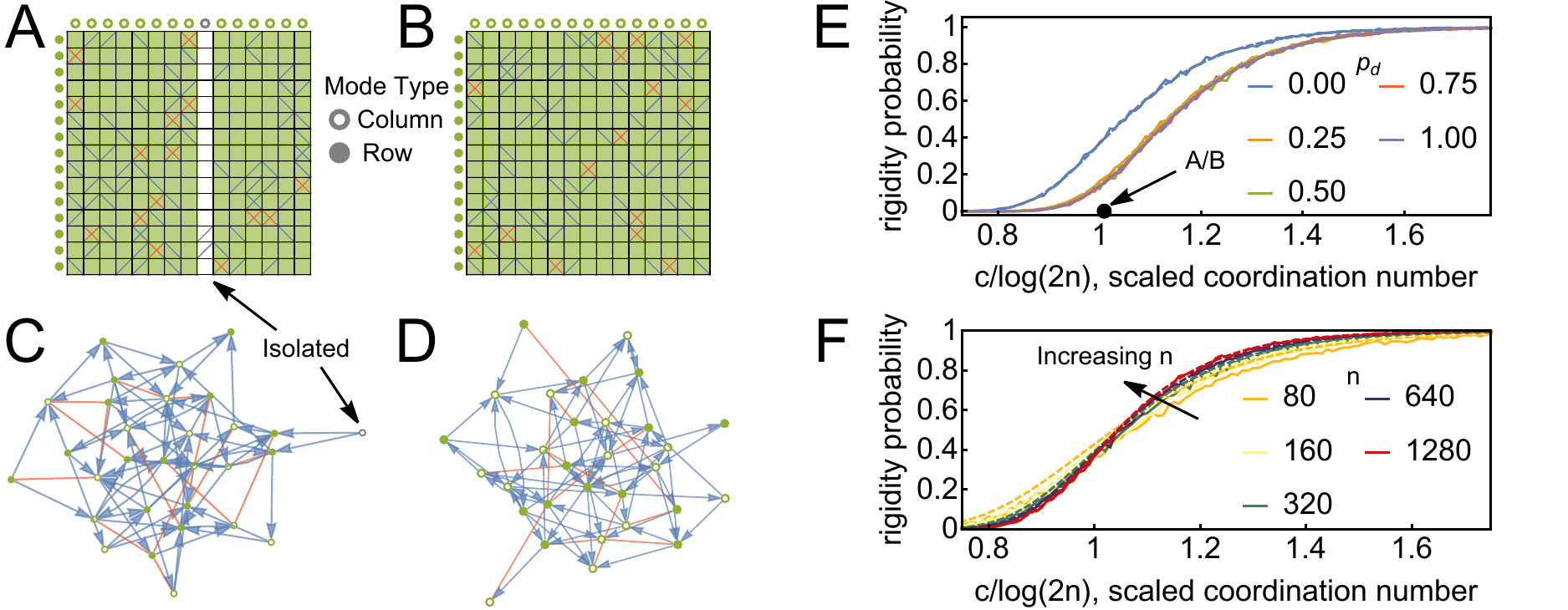}
    \caption{
    Complete rigidity, in which the last shearing mode is eliminated, is an entirely distinct process from the emergence of a rigid cluster.
    As additional elements are added, the lattice begins to saturate, such that only one or a few shearable columns or rows separate rigid (green) regions (A) until the lattice becomes fully rigid (B).
    (C) In the graph theory corresponding to the structure in (A), there can be only a single node that is not both upstream and downstream from the giant cluster, corresponding to the shearable column. In contrast, the graph in (D) corresponding to (B) is fully connected, therefore the structure  is fully rigid.
    As shown in (E), full rigidity emerges when the density of random elements per column is $O(\log n)$, in contrast with the $O(1)$ elements per column  seen in the giant cluster phase transition. The presence of any amount of cables ($\pd>0$) leads to a universal rigidity probability that is significantly lower than that of the system with purely rigid elements ($\pd=0$).
    (F) As system size increases, the slope of the rigidity probability transition increases approaching a step function in the thermodynamic limit. }
    \label{fig:fullRigidity}
\end{figure*}

A system with a single isolated mode, such as that shown in~\myfig{fullRigidity}(A) cannot be fully rigid, whereas a system such as that shown in~\myfig{fullRigidity}(B) is fully rigid despite not having a greater number of elements. In the graph theoretical analogs depicted in~\myfig{fullRigidity}(C,D), the single isolated vertex corresponds to a node in the former which is upstream only from the rest of the system.
As shown in Methods, in the thermodynamic limit, the requirement that no row or column mode be isolated is not only necessary but sufficient to ensure full rigidity, resulting in a probability:

\begin{equation}
\pr\approx\begin{cases}
          (1-e^{-\coord})^{2\n} \quad &\text{if} \, \pd = 0 \\
          (1-2e^{-\coord})^{2\n} \quad &\text{if} \, \pd \neq 0 \\
     \end{cases}
     \label{eq:rigidityCases}
\end{equation}

The first case ($\pd = 0$), in which only rigid rods are present, was previously derived via the graph-theoretic method~\cite{ellenbroek2011rigidity}.
Surprisingly, there is a new universal rigidity transition point when any amount of cables are present ($\pd \neq 0$). 
Note that this transition occurs around $c = \log(2n)$, in which the probability of a plaquette containing an added cable/rod is on the order of $\log(n)/n$, rather than on the order of $1/n$ at which the collective rigidity emerges via the giant cluster.
As shown in~\myfig{fullRigidity}(E), this analytic theory matches the simulation results. Per ~\myfig{fullRigidity}(F), the transition becomes gradually sharper as the system size increases, approaching a sharp transition to certain full rigidity at $c = \log(2n)$ in the infinite-size limit. 
However, because this occurs at logarithmic order, even thermodynamically large systems can have ranges of bond densities in which finite fractions of systems are fully rigid and others are not.

\begin{figure*}[t]
    \centering
    \includegraphics[width=2\columnwidth]{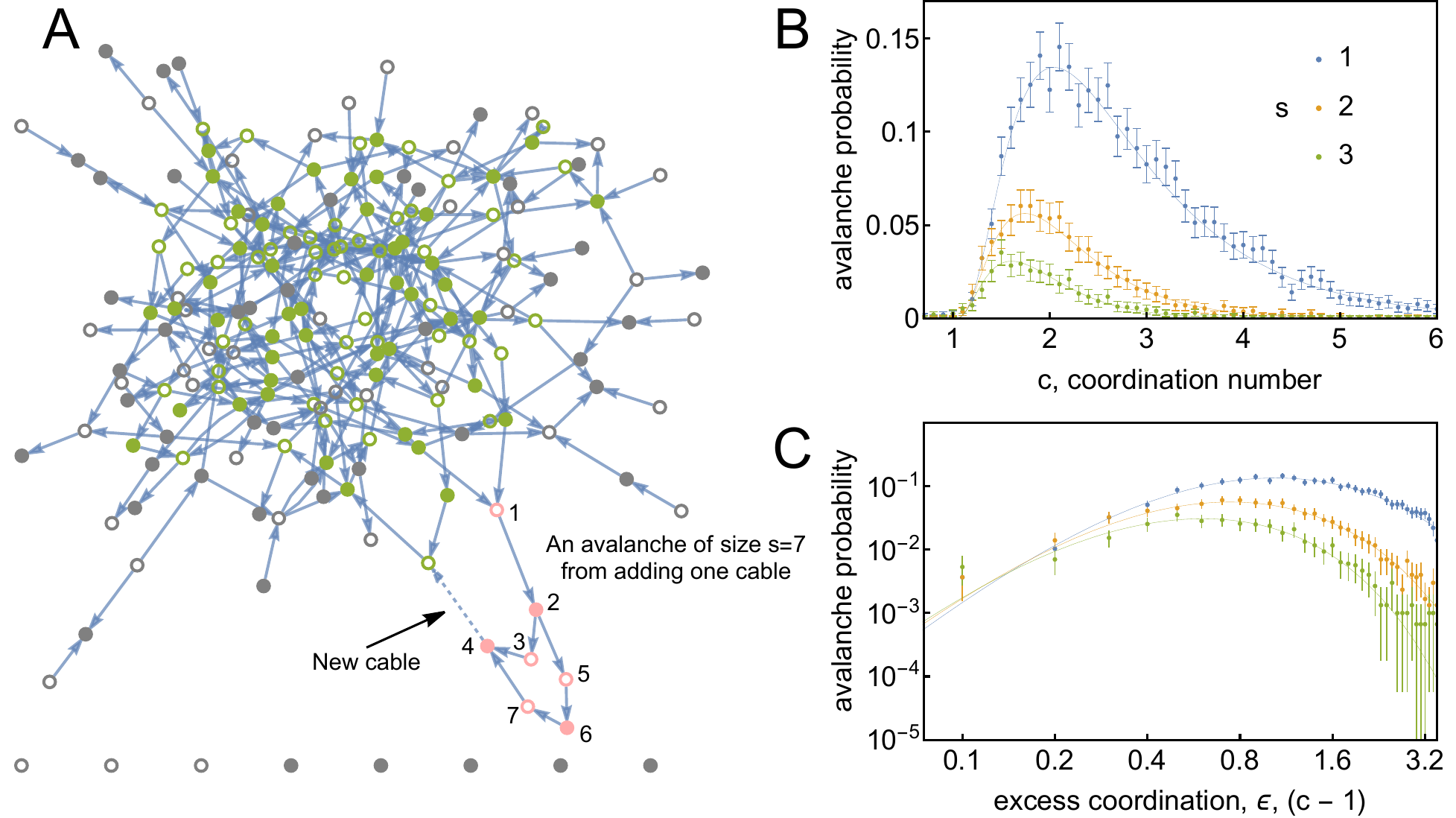}
    \caption{
    A hallmark of the tensegrity structure is an avalanche process, in which  multiple floppy modes are eliminated via a single constraint. 
    (A) The presence of directed edges (cables) in the graph means that groups of vertices such as the pink nodes can be downstream (or upstream) from the giant cluster without being a part of it.
    Consequently, because the pink nodes are all upstream from vertex 4, when a new (dashed) edge renders that vertex upstream from the giant cluster, all seven modes become eliminated at once in an avalanche.
   (B) Avalanches of increasing size $s$ occur with finite but decreasing frequency, in a manner largely consistent with an analytic mean-field theory (solid lines). 
   These avalanches are most frequent slightly above the transition at $c=1$, where the giant cluster grows most rapidly and decrease at larger coordination numbers for which large isolated groups become exponentially rare.
   (C) 
   Close to the transition, both theory and simulation are in agreement, showing power-law behavior.
    }
    \label{fig:avalancheFig}
\end{figure*}

\section{Avalanches}

Avalanches in non-equilibrium statistical systems, in which many degrees of freedom switch states at once, are quintessential signs of collective behavior, self-organized criticality~\cite{Wiesenfeld} and memory/history dependence~\cite{Sethna_2001}. However, while some mechanical systems (like jammed packings \cite{Henkes_2005,liu2010jamming} and fiber networks~\cite{broedersz2011criticality,Shivers_2019}) do display aspects of critical phenomena, the addition of a bond in linear constraint network necessarily either eliminates one zero mode or no zero modes, as has been known since the time of Maxwell~\cite{maxwell1864calculation}.

In contrast, in systems that contain cables and/or struts multiple zero modes can be eliminated by a single additional constraint, in an ``avalanche'' process. As shown in \myfig{avalancheFig}(A), avalanches occur when many independent modes that are already downstream (upstream) from the giant cluster are all made to be upstream (downstream) as well. In contrast, in systems with only rigid constraints, any two vertices joined by an undirected edge are always both upstream and downstream from each other, so no avalanches of size greater than one can occur. The ability to undergo large avalanches allows systems with cables to rapidly change behavior. Similar avalanche behavior can be seen in other network models, such as those that model outbreak sizes in disease spread~\cite{cai2015avalanche} or neural activity~\cite{beggs2003neuronal}.

We analytically predict the distribution of avalanches of size $s$ via a self-consistent mean-field theory that introduces the notion of a node being \emph{dependent} on another node.
We say that a vertex has $s$ upstream (downstream) dependents if an edge  added from (to) the node to (from) the giant cluster would result in an avalanche of size $s$ (note that each node upstream or downstream from, but not contained in, the giant cluster is its own dependent).
Looking again to \myfig{avalancheFig} (A), we see that vertex $4$ has seven (upstream) dependents ($s=7$), vertex $7$ has five dependents ($s=5$), and so on.

For a vertex with $k$ vertices directly upstream from it to have $s$ total upstream dependents, those $k$ vertices must have a total of $s-1$ upstream dependents (since this latter set does not include the original vertex). Thus, we can recursively solve for the fraction $D_s$ of vertices with any number $s$ dependents from the base case ($D_1$), as shown in Methods. 

Consider the placement of an additional bond, where for simplicity we assume that all bonds are directed. 
For an avalanche of size $s$ to occur, the new bond must either directly connect a vertex with $s$ (upstream or downstream) dependents to the giant cluster, or connect a vertex with $s_1$ upstream dependents to a vertex with $s_2$ downstream dependents such that $s_1+s_2=s$, leading to a probability:

\begin{align}
    p_{s} &= 2\fb D_{s} + \sum_{\substack{s_{1}+s_{2} = s\\s_{1},s_{2}>0}} D_{s_{1}}D_{s_{2}}.
\end{align}

As shown in \myfig{avalancheFig} (B), our theory is in agreement with simulation results. 
Agreement breaks down for larger avalanches and closer to the critical point, suggesting a breakdown of the mean-field theory, which presumes that a vertex in an avalanche is not a dependent of another in more than one way. For example, in \myfig{avalancheFig}~(A), vertex 2 is dependent on vertex 4 via two different paths.

As expected, only in the presence of a giant cluster ($\coord>1$) can avalanches occur. 
Immediately after the emergence of the giant cluster, where many more vertices are upstream from the cluster than in it, avalanches of any finite size are nearly equally likely. Indeed, in this regime, for avalanches small compared to the section of the graph upstream from the giant cluster, $p_s \approx (1+s)(2\exc)^4$, revealing that larger avalanches are \emph{more} likely than smaller avalanches, a trend that begins to emerge in \myfig{avalancheFig}~(C).

Consequently, the majority of the nodes that initially join the giant cluster do so through large avalanches. As the giant cluster grows and fewer independent modes remain, vertices with many dependents become exponentially less likely, resulting in large avalanches becoming more rare than small avalanches. Once nearly every vertex is in the giant cluster, even avalanches of size $s=1$ become rare, as there so few independent modes left to join the giant cluster.

\section{Discussion}

We have considered a model system with randomly placed cables to study the emergence of rigidity in systems involving inherently nonlinear elements such as cables, struts and membranes.
We demonstrate two phase transitions controlled by parameters that generalize Maxwell constraint counting. First, at bond density $c=1$, systems can support external forces through having extensive rigid area fractions. Second, at bond density $c=\log2n$, systems become fully rigid and can no longer self-deform. For both phase transitions, systems with mixtures involving \emph{any} amount of cables ($\pd \neq 0$) show distinctly different behavior than those with all rods ($\pd=0$). 

These systems also reveal fascinating non-equilibrium behavior, in which a structure can have multiple zero modes eliminated with the addition of a single constraint, in violation of the Maxwell-Calladine paradigm of linear constraints \cite{maxwell1864calculation}.
Indeed, immediately at the emergence of collective rigidity, larger avalanches are \emph{more} likely than smaller avalanches, though avalanche size gradually shrinks above the transition.
Similar avalanche behavior can be found in other systems ranging from sandpile models (self-organized criticality)~\cite{Wiesenfeld} to disease spread ~\cite{cai2015avalanche}.
Our avalanche model for the square lattice can not just explain tensegrity rigidity, but percolation in other systems where one-way interactions are possible.

Our analytic approach for finding tensegrity rigidity can help guide the engineering of tensegrity structures, such as towers \cite{song2022form}, bridges \cite{tensegrity_bridges}, domes \cite{tensegrity_domes}, or others \cite{Zhang2015}, as their demand continues to grow. It can also help our understanding of how complex biological tensegrities \cite{ingber2014tensegrity} arose through successive generations of random mutation and natural selection. It remains an open question how random tensegrities generate rigidity in more complicated structures, such as those lacking our regular square backbone or existing in three dimensions or with other types of nonlinear constraints such as tensioned elements or membranes.

\emph{Acknowledgements} The authors acknowledge helpful conversations with Xiaoming Mao. The authors gratefully acknowledge financial support from the Army Research Office through the MURI program (\# W911NF2210219).

\section*{Methods}

\subsection{Giant Cluster and Extensive Rigidity}

As described in the main text, the number of independent deformation modes of the random tensegrity structure is equal to the number of clusters (sets of nodes that are mutually reachable) in a graph with randomly placed directed and undirected bonds. To that end, we derive a mean-field theory, similar to those previously derived for directed \cite{karp1990transitive} and undirected \cite{newman2009random} graphs, calculating the number of clusters in a bipartite graph with $\n$ sites of each type, each with an average of $\cOne$ directed and $\cTwo$ undirected edges attached to it.

In particular, we are interested in a \emph{giant} cluster which contains an extensive (proportional to $\n$) number of nodes. It is possible for a system to contain zero such clusters or one such cluster, but multiple clusters occur vanishingly rarely in large systems, since the odds of them not being joined into a single cluster diminish with system size. To that end we refer to $\fb$, the fraction of nodes in the giant cluster and $\fu$ and $\fd$, the number of nodes that are upstream and downstream, respectively, from the cluster (note that each node in the cluster is both upstream and downstream from it). By symmetry $\fu = \fd$ and the number of nodes in the giant cluster is $2\fu-\fe$.

To determine these quantities, we make a mean-field assumption, in which the probabilities that two generic nodes are each in a cluster (or upstream/downstream from one) are independent from each other. Consider now the probability that a particular node is \emph{not} upstream from the giant cluster, $1-\fu$. This is exactly the probability that for each node the original node could have an edge to (not including directed edges in the wrong direction) that connection is absent \emph{or} the node in question is itself not upstream from the giant cluster. Under the mean-field assumption, the probabilities that nodes are upstream from the giant cluster are uncorrelated, and so the probability that any of the other nodes is not upstream from the giant cluster is also $1-\fu$. This introduces a self-consistency condition on $\fu$ that will allow us to determine it. Expressed mathematically, for a vertex $i$ to not be upstream from the giant cluster through a vertex $j$, this condition becomes:

\begin{align}
p&(i \textrm{ is not upstream through } j) = \nonumber\\
p&(i \textrm{ is not upstream from } j) ~+ \nonumber\\
p&(i \textrm { is upstream from }j\textrm{, but }j\textrm{ is not in } \fu) =\nonumber\\
&\left(1 - \frac{1}{n}\left(\frac{\cOne}{2} + \cTwo\right)\right) + \frac{1}{n}\left(\frac{\cOne}{2} + \cTwo\right)(1-\fu).
\end{align}

\noindent Of course, for vertex $i$ to not be upstream from the giant cluster, this condition must be true for all $\n$ possible choices of vertex $j$:

\begin{align}\label{eq:fuTranscendentalMethods}
1 - \fu &= \left[1 - \frac{1}{\n}\left(\frac{\cOne}{2} + \cTwo\right) + \frac{1}{\n}\left(\frac{\cOne}{2} + \cTwo\right)  (1-\fu)\right]^\n
\nonumber\\
&= \left[1 - \frac{1}{\n}  \left(\frac{\cOne}{2} + \cTwo\right) \fu\right]^\n
\nonumber\\
&\rightarrow e^{-\left(\cOne/2 + \cTwo\right)  \fu} \equiv e^{-\coord \fu},
\end{align}

\noindent where the final expression is in the thermodynamic limit and our definition for $\coord \equiv \cOne/2 + \cTwo$ is the average number of outgoing bonds per site.

Because our giant cluster statistics are determined by two distinct quantities, $\fu, \fe$ we must generate a similar expression for self-consistency involving the latter. To that end, we now calculate the probability $1-\fe$ that a given node is neither upstream nor downstream from the giant cluster. In considering the node's potential connection to another node, we must now consider four scenarios, an undirected edge, an outgoing directed edge, an incoming directed edge, or no edge. In order for the original node to be neither upstream or downstream, in those four cases, we require that the second node be, respectively, neither upstream nor downstream, not be upstream, not be downstream, or have any arbitrary relation to the giant cluster. Expressing again this relationship in mathematical terms for a vertex $i$ to not be upstream or downstream from the giant cluster through a vertex $j$, and again making the mean-field and large-size assumptions, we obtain:

\begin{align}
p&(i \textrm{ is not upstream or downstream through } j) = \nonumber\\
p&(i \textrm{ has no bonds with } j) ~+ \nonumber\\
p&(i \textrm { has a directed edge to }j\textrm{, but }j\textrm{ is not in } \fu) =\nonumber\\
p&(i \textrm { has a directed edge from }j\textrm{, but }j\textrm{ is not in } \fd) =\nonumber\\
p&(i \textrm { has an undirected edge with }j\textrm{, but }j\textrm{ is not in } \fe) =\nonumber\\
&\left(1 - \frac{1}{n}\left(\cOne + \cTwo\right)\right) + \frac{1}{n}\left(\frac{\cOne}{2}(1-\fu)\right)+ \nonumber\\ 
&\frac{1}{n}\left(\frac{\cOne}{2}(1-\fu)\right)+\frac{1}{n}\left(\cTwo(1-\fe)\right) = \nonumber\\
&= 1-\frac{1}{n}\left(\cOne \fu + \cTwo \fe \right),
\end{align}

\noindent Again, for vertex $i$ to not be upstream or downstream from the giant cluster, this condition must be true for all $n$ possible choices of vertex $j$:

\begin{align}\label{eq:feTranscendentalMethods}
1 - \fe &= \left[1-\frac{1}{n}\left(\cOne \fu + \cTwo \fe \right)\right]^n \\ \nonumber
&\rightarrow e^{-\left(\cOne \fu + \cTwo \fe \right)}.
\end{align}

\begin{figure*}[t]
    \centering
    \includegraphics[width=0.95\textwidth]{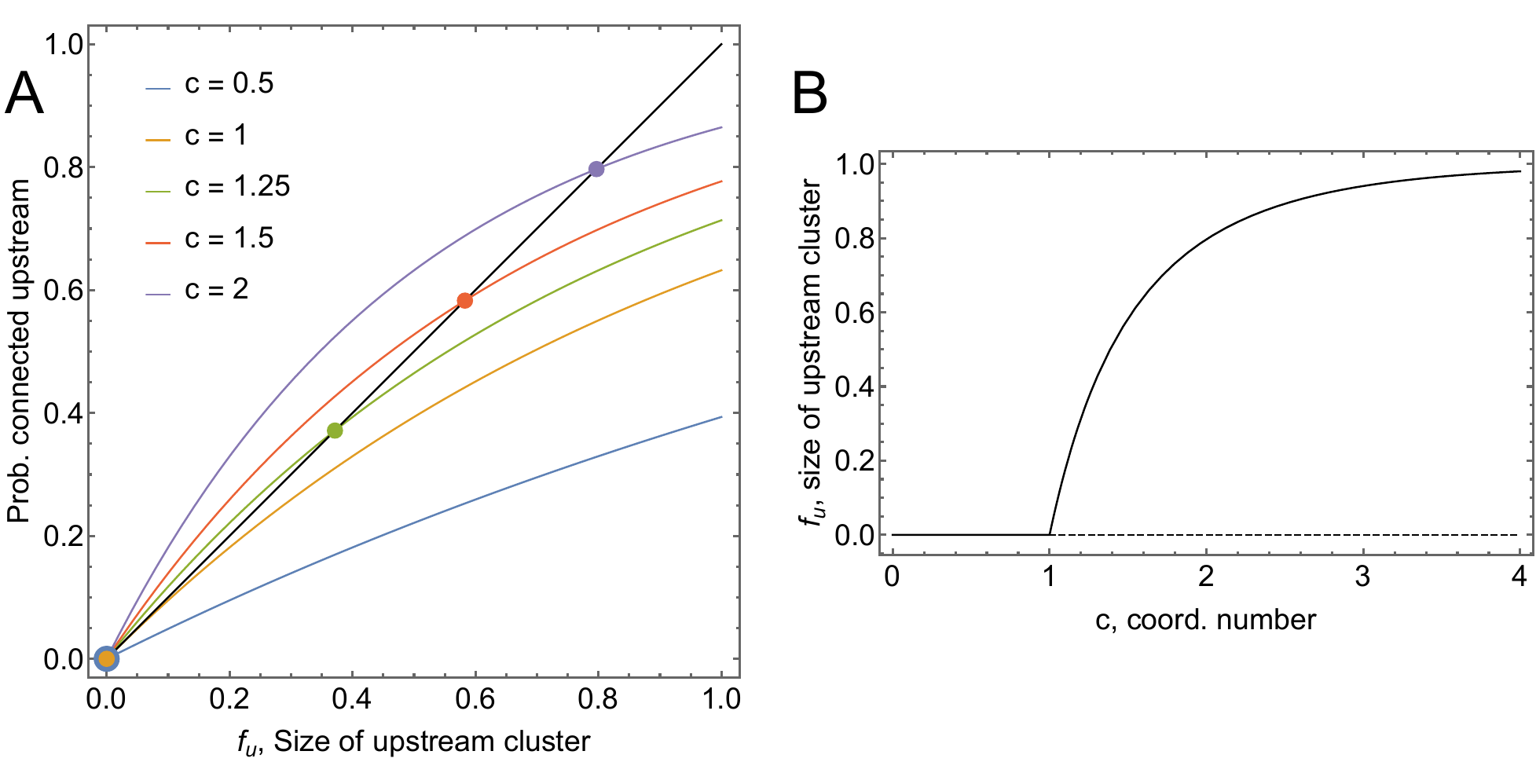}
    \caption{(A) Self-consistency requires that the fraction of nodes in the upstream cluster (black line) correspond to the probability that a given node is upstream from the giant cluster (colored lines), as calculated in subsection A of Methods. This latter quantity depends on $\coord$, the average number of outgoing edges emanating from a node. The marked transcendental intersection points yield the fraction $\fu$ of nodes upstream from the giant cluster. This quantity is plotted in (B), with the dashed line indicating that even beyond the phase transition at $\coord=1$ an upstream cluster of size zero satisfies the self-consistency equations.}
    \label{fig:meanfield2}
\end{figure*}

We have now obtained two transcendental equations (\myeq{fuTranscendentalMethods} and \myeq{feTranscendentalMethods}) in two variables, to which no analytic solution exists, other than the trivial $\fu = \fe = 0$. However, such a nontrivial solution begins to exist at $\coord > 1$, and can be obtained numerically as shown in \myfig{meanfield2}. The transition point for this nontrivial solution ($\coord=1$) can be found analytically, through finding when the two sides of \myeq{fuTranscendentalMethods} plotted in \myfig{meanfield2} are tangent to each other. Subsequently, $\fu$ is used as a parameter in \myeq{feTranscendentalMethods} which is likewise solved to obtain $\fe$. These mean-field solutions are validated by simulations performed in Mathematica representing systems of size up to $10^6$ pores.

To determine the critical behavior of the square-lattice system just above its critical point at $\coord=1$, we Taylor expand the transcendental equations  for $\fu$ and $\fe$ (\myeq{fuTranscendentalMethods} and \myeq{feTranscendentalMethods}) around $\exc \equiv \coord - 1$ and solve for $\fu$ and $\fe$:

\begin{align}\label{eq:expandedFu}
\fu &= 2\exc -\frac{8}{3}\exc^2\\
\label{eq:expandedFe}
\fe &= 4\exc - \left(\frac{4}{\pd}+\frac{16}{3}\right)\exc^2.
\end{align}

\myeq{expandedFe} is undefined when $\pd=0$. However, in the case of $\pd=0$ where all edges are undirected, there is no way for a vertex to be upstream but not downstream from the giant cluster, giving $\fe=\fu$. With these two distinct cases for $\fe$, we can solve for two distinct cases of $\fb$ using the inclusion-exlusion relation $\fb=2\fu-\fe$:

\begin{align}\label{eq:fblimitcases}
\fb\approx\begin{cases}
          \frac{4}{\pd}\exc^2 \quad &\text{if} \, \pd \neq 0 \\
          2\exc-\frac{8}{3}\exc^2 \quad &\text{if} \, \pd = 0 \\
     \end{cases}
\end{align}

\noindent In the presence of any cables ($\pd>0$) there is quadratic growth of the giant cluster and therefore quartic growth of the rigid area fraction. In the unique case of all rods ($\pd=0$), the giant cluster grows linearly resulting in quadratic rigid area fraction growth (see \myfig{rigidAreaFrac}). These two cases are unusual in that the case for $\pd \neq 0$ does not aproach the case of $\pd=0$ as $\pd \rightarrow 0$, they are completely different behaviors. This is not too surprising mathematically, as in the derivation for $\fe$ in the general $\pd \neq 0$ case we divide by $\pd$, making any solution completely nonphysical when $\pd=0$.

\subsection{Full Rigidity}

Full rigidity, in which the square lattice system can no longer self-deform and is constrained to its reference square shape, is determined when every row and column vertex belongs to a single strongly connected cluster. Equivalently, for every vertex to belong to the giant cluster, there must be no independent sets of $m$ nodes (including single vertices, which are sets of size one) that are not strongly connected to the rest of the system (here, we say that a set, which is not necessarily itself a connected cluster, is strongly connected to another set if it contains elements that are upstream and elements that are downstream from elements in the second set).
Here, we derive the probability that no such independent sets exist, which reveals that the addition of any amount of cables to a system of rigid elements qualitatively changes the nature of the transition to full rigidity.
Our approach is to consider a set of $m \ll 2n$ nodes and calculate the probability that they are \emph{not} fully connected to the large, $(2n-m) \approx 2n$-node set that comprises the rest of the system. Ultimately, we will see that in the thermodynamic limit the case $m=1$ dominates.
To obtain this probability, we find both the probability that this small set does not connect to the large set through undirected edges (similar to the approach of~\cite{ellenbroek2011rigidity}) and the probability that it does not have directed edges both to and from the large set.
 
For a generic set of $m$ nodes to be independent of the rest of the system, there must be no undirected edges that connect from any of the $m$ vertices in the finite set to any of the approximately $\approx n$ vertices of the other type (of the bipartite graph) in the large set:

\begin{align}\label{eq:undirectedisolation}
    p_{\textrm{undirected}}^{\textrm{iso}}(m) &= \left[\left(1-\frac{\cTwo}{\n}\right)^\n\right]^\m
    \nonumber\\
    &\approx \left[e^{-\cTwo}\right]^\m,
\end{align}

\noindent where the approximation is valid in the thermodynamic limit of $n \rightarrow \infty$.

A single directed edge connecting the small set of nodes to the rest of the system would only make it downstream or upstream from the giant cluster, and the system would still be guaranteed to possess multiple deformation modes. 
However, if there are two directed edges that make the small set both upstream and downstream from the rest of the system, it would no longer be isolated and no longer necessarily represent an independent deformation mode. Consequently, for a small set to not be connected via directed edges, it is allowed have either an upstream or downstream connection to the rest of the system, but not both. The probability that it \emph{is} both upstream and downstream is the square of the probability that there is a directed edge to/from the giant cluster:

\begin{align}
p_{\textrm{directed}}^{\textrm{iso}}(m) &\approx 1-\left( 1-\left[e^{-\cOne/2}\right]^\m \right)^2\nonumber\\
   &= \left[e^{-\cOne/2}\right]^\m\left(2-\left[e^{-\cOne/2}\right]^\m\right).
\end{align}

Of course, for a set to be isolated, it must be isolated by both undirected and directed edges. As the probabilities for undirected isolation and directed isolation are independent of each other, the probability of the cluster being isolated is their product:

\begin{align}
   p^{\textrm{iso}}(m) &= p_{\textrm{undirected}}^{\textrm{iso}}(m)
   \,  p_{\textrm{directed}}^{\textrm{iso}}(m)
   \nonumber
   \\
   &\approx \left[e^{-\left(\cTwo + \cOne/2\right)}\right]^\m\left(2-\left[e^{-\cOne/2}\right]^\m\right)\nonumber\\
   &\approx \left[e^{-\coord}\right]^\m\left(2-\left[e^{-\cOne/2}\right]^\m\right).
\end{align}

In the thermodynamic limit, the probability of isolated sets of $m=1$ dominates over larger sets. Specifically, when there are sufficient bonds that there is nontrivial probability that there are no isolated single nodes, the probability that there are isolated larger sets without there being isolated single nodes becomes negligible. Thus, the probability of full rigidity, is the probability that no single node is isolated from the rest of the system:

\begin{align}
   \pr &\approx \left(1-\pIso(1)\right)^{2n}\nonumber\\
   &\approx\left(1-\left(2-e^{-\cOne/2}\right)e^{-\coord}\right)^{2n}.
\end{align}

From the above expression, in order for there to be nonnegligible probability of rigidity, we need $c$ to be on the order of $\log n$. Hence, in contrast with the emergence of the giant cluster at $c = 1$, corresponding to an average of one rigid rod (or two cables) per column of $n$ squares, full rigidity occurs when the density of rods per square is on the order of $(\log n)/n$.

In the generic case where $\cOne$ is proportional to $\coord$ ($\pd\neq0$), $\cOne$ would also be very large, resulting in $(e^{-\coord})(e^{-\cOne/2})$ becoming vanishingly small. However, when $\pd=0$, $\cOne=0$, making $e^{-\cOne/2}=1$. These two cases result in two distinct coefficients in $\pIso$, giving two distinct rigidity probabilities:

\begin{equation}\label{eq:prCasesMethods}
\pr\approx\begin{cases}
        (1-2e^{-\coord})^{2\n} \quad &\text{if} \, \pd \neq 0 \\
        (1-e^{-\coord})^{2\n} \quad &\text{if} \, \pd = 0.\\
     \end{cases}
\end{equation}

As the case for $\pd\neq0$ has a smaller base being raised to the $2\n$, the probability of rigidity in the presence of any extensive number of cables will always be less than the probability of rigidity in the unique case of all rods. Furthermore, we see that unlike in the transition to extensive rigidity, the probability of full rigidity is dependent on system size. 

To examine the difference in behavior of these two probabilities as the system size changes, let the coefficient on $e^{-c}$ be denoted by $\sigma$, where $\sigma=2$ when $\pd\neq0$ and $\sigma=1$ when $\pd=0$. We now solve for the scaled coordination number $\coord$ that corresponds to a given $\pr$ in terms of $\sigma$:

\begin{align}
    (1-\sigma e^{-\coord})^{2\n} &\approx \pr \nonumber\\
    2n\sigma e^{-c} &\approx -\log{\pr} \nonumber\\ 
    -\coord + \log{2n\sigma} &\approx \log{(-\log{\pr})} \nonumber\\
    \coord &\approx \log{2n}-\log{(-\log{\pr}}) +\log{\sigma}
\end{align}

We see that the coordination number for a given probability has a base term independent of $\pr$ and $\sigma$: $\log{2\n}$. In order to account for this term's dependence on system size, we divide through by $\log{2\n}$, giving the \emph{scaled} coordination number for a given $\pr$:

\begin{align}
    \frac{\coord}{\log{2n}} &\approx 1-\frac{\log{(-\log{\pr}})}{\log{2n}} +\frac{\log{\sigma}}{\log{2n}}
\end{align}

The third term on the right-hand side is the only dependence on $\sigma$, representing the difference between the two cases in \myeq{prCasesMethods}. As this term is divided by the logarithm of the system size, in the true thermodynamic limit it vanishes and the two cases unify. However, as it is the logarithm of the system size that must be large for this term to vanish, the cases being distinct is an incredibly persistent finite-size effect. 

The second term on the right-hand side, which is the only dependence on $\pr$, is also divided by the logarithm of the system size. Therefore, in the thermodynamic, in order for $\pr$ to not be close to either 0 or 1, $\coord$ must be approximately $\log 2\n$. Physically, this indicates that in this limit for $c>\log 2 n$ the system will be rigid with high probability and for $c < \log 2 n$ the system will not, indicating that the probability of rigidity (slowly) approaches a step function of the bond concentration.

\begin{figure}
    \centering
    \includegraphics[width=0.45\textwidth]{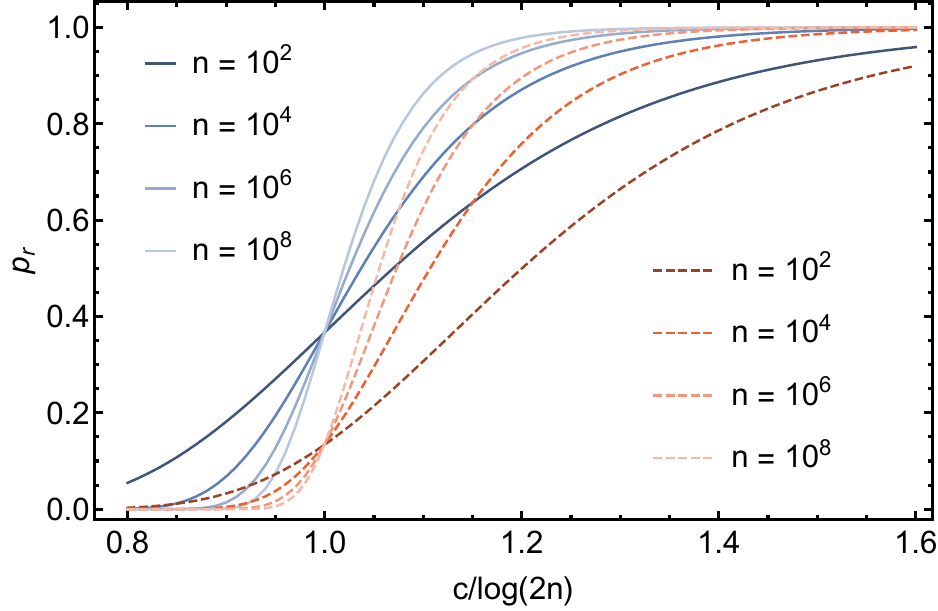}
    \caption{The probability of full rigidity when $\pd=0$ (blue, solid) and $\pd \neq 0$ (red, dashed). As system size grows exponentially (as the color of the lines gets lighter), both rigidity cases approach a step function centered at $\coord=\textrm{Log}[2n]$ and the difference between the two cases decreases.}
    \label{fig:prSystemSize}
\end{figure}

\subsection{Avalanches}

As discussed in the main text, as cables are randomly added to a structure of rigid rods (for simplicity, we do not include randomly placed rods in our avalanche analysis), the addition of a single cable can eliminate multiple zero-energy deformation modes. We refer to the elimination of $s$ such zero modes as an \emph{avalanche} of size $s$.
Here, we derive the probability of these avalanches based on the statistics of how nodes in the associated graph theory join the giant cluster.
In pursuit of this goal, we introduce the notion of \emph{dependents}, nodes that automatically join the giant cluster when the vertex they are dependent on joins the giant cluster.
Let $D_{s}$ represent the fraction of vertices that have $s$  dependents (this is defined in either the upstream or downstream direction, but the two quantities are equal by symmetry), including itself. 

Two scenarios can result in an avalanche of size $s$.
First, a new bond may connect a vertex with $s$ dependents directly to the giant cluster. Second, a new bond may connect a vertex with $s_1$ upstream dependents to a vertex with $s_2$ downstream dependents such that $ s_1 + s_2 = s$. This implies that the probability of such an avalanche is

\begin{align} \label{eq:probability_avalanche}
    p_{s} &= 2\fb D_{s} + \sum_{\substack{s_{1}+s_{2} = s\\s_{1},s_{2}>0}} D_{s_{1}}D_{s_{2}},
\end{align}

\noindent where $\fb$ is the fraction of nodes in the giant cluster. 
The first term is the probability of a vertex with $s$ dependents connecting directly to the giant cluster, where the factor of two accounts for dependents being either upstream and downstream. The second term sums over all possible ways for $s$ total dependents to exist between an upstream and downstream pair of vertices.

Consider a vertex with only itself as an (upstream) dependent.
It must then be directly connected to the giant cluster without having any other dependents, requiring that it:

\begin{enumerate}
  \item have at least one directed edge that connects directly to the giant cluster,
  \item not be downstream from the giant cluster, as it would then be contained in the giant cluster and
  \item not be upstream from a vertex that is upstream, but not contained in, the giant cluster, as 
  that would result in having more than one dependent.
\end{enumerate}

\noindent From these three conditions we can say that the fraction of nodes with exactly one dependent is

\begin{align}
    D_1 = (1-e^{-\cOne\fb/2})(1-\fu)(e^{-\cOne(\fe-\fu)/2}).
\end{align}

\noindent Where $\fu$ is the fraction of nodes upstream from the giant cluster and $\fe$ is the fraction of nodes either upstream or downstream from the giant cluster.

For a vertex to have $s>1$ (upstream) dependents, it must have at least one ``direct dependent'', a dependent that it directly connects to. 
Such a direct dependent must necessarily be among the $n(\fe-\fu)$ vertices that it could attach to that are upstream from but not contained in the giant cluster.
Let $0<k<s$ denote the number of direct dependents of a particular vertex.
The probability of connecting to exactly $k$ upstream-only nodes (regardless of whether they are in fact dependents) is simply a binomial problem:

\begin{align}\label{eq:k}
    p_k&=\binom{n(\fe-\fu)}{k} \left(\frac{\cOne}{2n}\right)^k \left(1-\frac{\cOne}{2n}\right)^{n(\fe-\fu)-k}\nonumber\\
    &\approx \left(\frac{\cOne(\fe-\fu)}{2}\right)^k \frac{1}{k!} e^{-\cOne(\fe-\fu)/2}.
\end{align}

Then, for this vertex with $k$ direct dependents to have $s$ total dependents, the dependents of the $k$ direct dependents must sum to $s-1$ (because they include all of the dependents of the original node except for itself). 
We take the probability of having $k$ dependents and multiply it by the possible conditional probabilities that each direct dependent has $s_j$ dependents, where the sum of all of the $s_j$ must be $s-1$. Importantly, for this vertex to have dependents, it must not be downstream from the giant cluster, giving a factor of $1-\fu$:

\begin{align} \label{eq:dependence_probability}
    D_{s} = (1-\fu)\sum_{k=1}^{s-1} p_{k}\sum'\left(\prod_{j=1}^{k}\frac{D_{s_{j}}}{\fe-\fu}\right),
\end{align}
\noindent where $\sum'$ denotes a sum over all sets of avalanche sizes with each $s_j>0$ and $\sum_j s_j = s-1$.

\myeq{dependence_probability} can be simplified by substituting in \myeq{k} and factoring out terms from the summation and product notations:

\begin{align}
    D_{s} = (1-\fu)e^{-\cOne(\fe-\fu)/2}\sum_{k=1}^{s-1} \left(\frac{\cOne}{2}\right)^k\frac{1}{k!} \sum'\left(\prod_{j=1}^{k}D_{s_{j}}\right).
\end{align}

\noindent While the coefficients $D_s$ are physically meaningful, the above expression implies a relationship to the bond density $c/2$ that is not actually connected to the combinatorial aspects of the problem. To separate these two aspects of the problem, we define the reduced dependent fraction

\begin{align}\label{eq:dsDef}
d_s \equiv \left((1-\fu)e^{-\cOne(\fe-\fu)/2}\right)^{-s}  \left(\frac{\cOne}{2}\right)^{-(s-1)}.
\end{align}

This leads to a recursion relation with purely combinatorial factors:

\begin{align}\label{eq:dsRecursion}
    d_{s} = \sum_{k=1}^{s-1} \frac{1}{k!} \sum'\left(\prod_{j=1}^{k}d_{s_{j}}\right),
\end{align}

\noindent with a base case of

\begin{align}\label{eq:d1}
    d_{1} = 1-e^{-\fb\cOne/2}.
\end{align}

From the recursion relation \myeq{dsRecursion} and inverting \myeq{dsDef} we can swiftly generate the probabilities of avalanches of moderate size as a function of the bond concentration $c_1/2$, as used to produce the predictions in \myfig{avalancheFig} of the main text.

In addition to the general case, there is an important limit of $\exc=c-1 \ll 1$, right above the transition at which the giant cluster emerges and avalanches become possible. Using the form of the giant cluster size derived in Methods Sec.~A [\myeq{fblimitcases}], in this limit \myeq{d1} becomes:

\begin{align}\label{eq:d1Approx}
    d_{1} \approx 4\exc^2.
\end{align}

 The recursion relation of \myeq{dsRecursion} shows that in this limit $d_{s>1}$ avalanches involving nodes with multiple direct dependents are higher-order in $\exc$. Thus, for example, an avalanche of size three is overwhelmingly more likely to involve a node connected to another node connected to a third, rather than having the first node directly connected to both of the other two. Consequently, the sum over all possible numbers of direct dependents $k$ in \myeq{dsRecursion} becomes:

\begin{align}\label{eq:dsApproxRecursion}
    d_{s} &\approx d_{s-1} \rightarrow\nonumber
    \\
    d_s &\approx 4\exc^2.
\end{align}

\noindent \myeq{dsDef} in the small-$\exc$ limit gives $D_s=d_s$, which, substituted into \myeq{probability_avalanche}, gives:

\begin{align}\label{eq:psApprox}
    p_s=16\exc^4(1+s).
\end{align}

\noindent Surprisingly, at this critical transition point, larger avalanches are \emph{more} likely to occur than smaller ones. As shown in simulation results presented in the main text (\myfig{avalancheFig}), the dominant frequency of the smallest avalanches indeed vanishes close to the transition point.

\begin{figure}[t]
    \centering
    \includegraphics[width=0.45\textwidth]{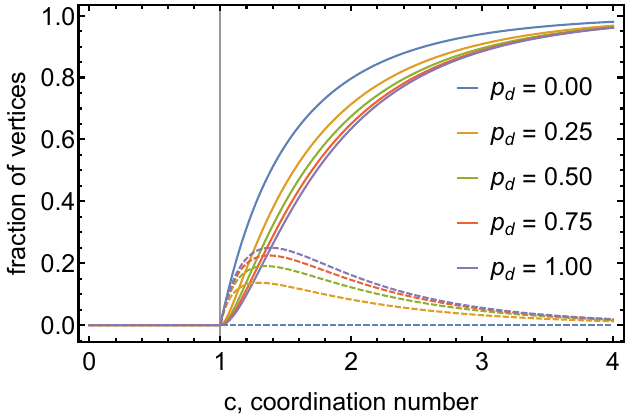}
    \caption{
    While at most coordination numbers the fraction of vertices in the giant cluster (solid) is much larger than the fraction of vertices upstream from but not contained in the giant cluster (dashed), this is not true in the regime just above the transition (for all $\pd\neq0$). Here, the fraction of ``upstream-only'' vertices initially grows rapidly, but peaks early as fewer vertices are able to be upstream but not contained in a growing giant cluster as $\coord$ increases.
    }
    \label{fig:fbfuonly}
\end{figure}

It might seem counter-intuitive that such large avalanches peak in probability when the giant cluster is so small and unlikely to connect to, such as $s=3$ in \myfig{avalancheFig}. For a vertex to have dependents at all, it must have at least one path through which either it is reachable from or can reach the giant cluster. However, for large avalanches, each vertex that joins the giant cluster must have first been upstream (or downstream) from the giant cluster but not contained in it. So while the probability of the upstream/downstream group being anchored to the giant cluster is proportional to $\fb$, there is a factor of $\fe-\fu$ (fraction of vertices ``upstream only'' or ``downstream only'' from the giant cluster) for \emph{each} vertex in the group. As shown in \myfig{fbfuonly}, this fraction of ``upstream only'' vertices initially grows much faster than the giant cluster itself, and peaks shortly after the transition, explaining why large avalanches are initially common and rapidly decrease in frequency.

In the regime just after the transition, it is particularly apparent that the fraction of vertices ``upstream only'' outpaces the giant cluster. When $\pd \neq 0$, $\fu \propto \exc$ (\myeq{expandedFu}), while $\fb \propto \exc^2$ (\myeq{fblimitcases}). Consequently, the fraction of vertices upstream from but not contained in the giant cluster is large compared to the vertices in the giant cluster right above the transition. Having so many vertices ``upstream-only'' from the giant cluster without many vertices in the giant cluster for upstream-only groups to be anchored to hints that there are few large groups of upstream-only vertices, helping to explain the diverging size scale of avalanches in this regime.

\begin{figure}[t]
    \centering
    \includegraphics[width=0.5\textwidth]{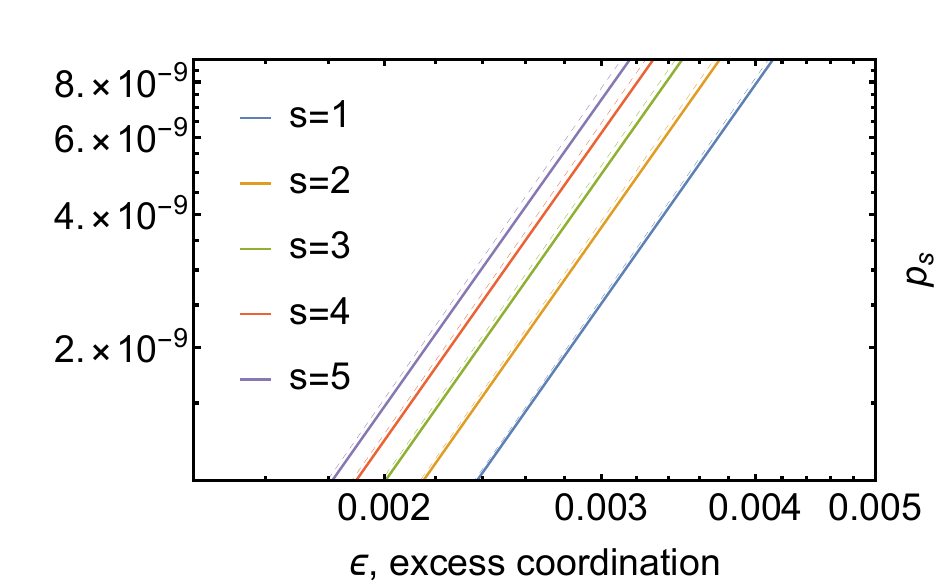}
    \caption{In the limit of small excess coordination ($\exc = \coord-1 \gtrsim 0$), the probability of an avalanche of size $s$ approximately follows $16\exc^4(s+1)$ (dashed). This approximation matches well with numerical solutions to the theoretical prediction of avalanches in the small excess coordination regime (solid).}
    \label{fig:psCritical}
\end{figure}

While we are unable to get simulation data for avalanches at really small excess coordination numbers (due to the many trials needed to determine small probabilities and the large sample sizes needed for precise coordination values), we nonetheless show that the small excess coordination approximation for avalanche probability from \myeq{psApprox} matches with the numerical solution to \myeq{probability_avalanche} in \myfig{psCritical}.

\bibliography{references.bib}

\end{document}

%% file: commandsNotation.tex
\newcommand{\shear}{\theta}
\newcommand{\row}{r}
\newcommand{\col}{c}

\newcommand{\n}{n} 
\newcommand{\coord}{c} 
\newcommand{\cOne}{c_1} 
\newcommand{\cTwo}{c_2} 

\newcommand{\exc}{\varepsilon} 

\newcommand{\fu}{f_u} 
\newcommand{\fd}{f_d} 
\newcommand{\fb}{f_b} 
\newcommand{\fe}{f_e} 
\newcommand{\vertexi}{i} 
\newcommand{\vertexj}{j} 
\newcommand{\m}{m} 

\newcommand{\pd}{p_d} 
\newcommand{\pr}{p_r} 
\newcommand{\pIso}{p^{\textrm{iso}}} 
\newcommand\myeq[1]{Eq.~(\ref{eq:#1})}
\newcommand\myfig[1]{Fig.~(\ref{fig:#1})}